\pdfoutput=1
\documentclass[twocolumn,preprintnumbers,nofootinbib,prl]{revtex4}

\usepackage{amssymb,amsmath,graphicx}
\usepackage{epsf}
\usepackage{graphicx,epsfig}
\usepackage{amsfonts}
\usepackage{amssymb}

\def\bk{{\bf k}}
\def\bp{{\bf p}}

\def\bx{{\bf x}}

\def\CH{{\cal H}}
\def\CL{{\cal L}}
\def\CO{{\cal O}}

\def\mpl{M_{\rm Pl}}
\def\half{\frac{1}{2}}

\begin{document}

\newcommand{\be}{\begin{equation}}
\newcommand{\ee}{\end{equation}}
\newcommand{\bea}{\begin{eqnarray}}
\newcommand{\eea}{\end{eqnarray}}
\newcommand{\barr}{\begin{array}}
\newcommand{\earr}{\end{array}}

\pagestyle{plain}


\title{Anomalous High Energy Dependence in Inflationary Density Perturbations}

\author{Xingang Chen$^{1}$ and Yi Wang$^{2}$}

\affiliation{
$^1$Centre for Theoretical Cosmology, DAMTP, University of Cambridge, Cambridge CB3 0WA, UK\\
$^2$Physics Department, McGill University, Montreal, H3A2T8, Canada
}


\begin{abstract}
We study the contribution of spectator massive scalar fields to the inflationary density perturbations through the universal gravitational coupling. We find that such contribution has several remarkable properties: it does not decrease as the mass of the spectator field increases; it has a significant size and cannot be turned off by any adjustable parameters; and it applies to all massive scalars existed during inflation, making the overall effect unexpectedly large. As a result, the primordial density perturbations are anomalously sensitive to the high energy physics.
\end{abstract}
\maketitle

Fields with masses much larger than the Hubble parameter are abundant in any realistic models of inflation. The usual description of inflation models are given in terms of one or several light degrees of freedom in terms of the low energy effective field theory, such as done since the earliest models \cite{Guth:1980zm,Mukhanov:1981xt}. The implicit assumption is that the heavier fields are effectively decoupled, analogous to the situation in particle physics.

For example, although the massive fields can be coupled to the inflaton through some model-dependent direct couplings in the matter Lagrangian, it has been shown through examples that their contribution to inflationary density perturbations is inversely proportional to some power of mass \cite{Kaloper:2002uj,Chen:2012ge}. Therefore heavier fields are less important, consistent with the intuition.
Of course, certain low-lying massive fields may become more accessible if the model contains sharp features \cite{Chen:2011zf,Achucarro:2010da} or special initial conditions \cite{Burgess:2002ub} that introduce higher energy scales, analogous to the injection of energies in particle physics colliders. But this does not affect the conclusion for the much heavier fields.

\begin{figure}
\begin{center}
\includegraphics[width=6cm]{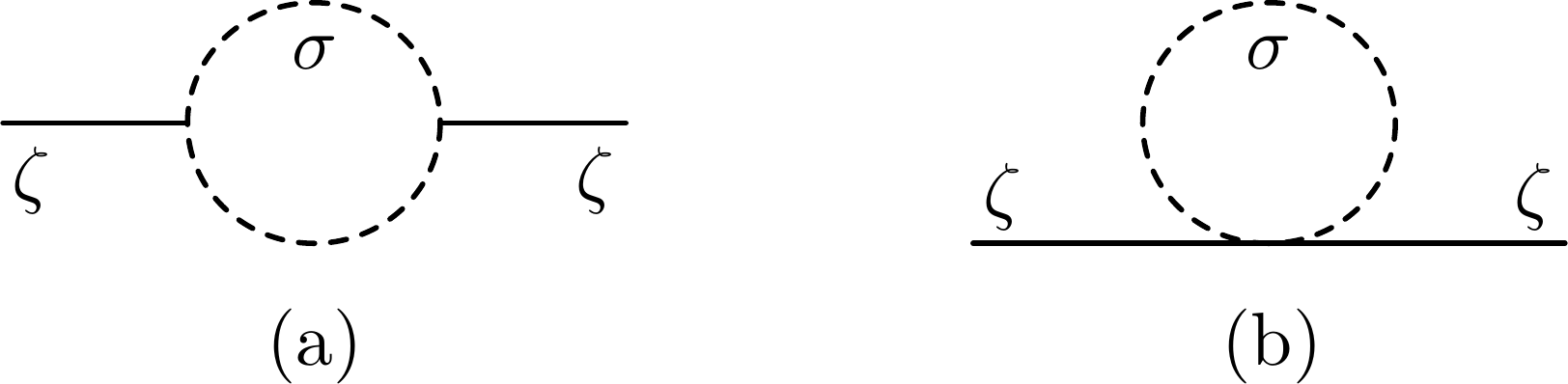}
\vspace{-0.6cm}
\end{center}
\caption{Two one-loop diagrams. $\zeta$ is the scalar perturbation, $\sigma$ is the massive field.}
\label{Fig:Loop_Diagrams}
\end{figure}

While the direct couplings may be present rather generically and exhibit many interesting phenomenologies, they are not the minimal case. Namely, their strength is parameterized by certain coupling parameters, and there always exists a model-building limit in which we can turn them off. However, for inflaton after we turned off all the direct couplings, there is still one coupling remaining -- the gravitational coupling -- which couples {\em all} fields existed during inflation to the inflaton and forms the baseline for all inflationary models.
This is the subject of this paper. We would like to compute the contribution of the massive fields to the inflationary density perturbations through the gravitational coupling. At one-loop level, this is given by the two diagrams in Fig.~\ref{Fig:Loop_Diagrams}. Due to the special properties of gravity, we will see many dramatic differences from the other cases.

We start with one massive scalar $\sigma$ with mass $M$ in a slow-roll inflation model. The matter Lagrangian is
\begin{align}
\CL_m = \sqrt{-g} \left[ -\half (\partial_\mu \phi)^2 - V_{\rm sr}(\phi)
- \half (\partial_\mu \sigma)^2 - \half M^2 \sigma^2 \right] ~.
\label{L_matter}
\end{align}
The $\phi$ is the inflaton and $V_{\rm sr}(\phi)$ is the slow-roll potential, with slow-roll parameter $\epsilon \equiv -\dot H/H^2$.
The massive field $\sigma$ is always decoupled from the inflaton in the matter Lagrangian and classically stays at $\sigma_0(t)=0$.

To study the density perturbations in the in-in formalism \cite{Weinberg:2005vy,Chen:2010xka}, we use the ADM formalism \cite{Arnowitt:1962hi,Maldacena:2002vr},
\bea
ds^2=-\hat N^2 d\hat t^2 + \hat h_{ij} (d\hat x^i + \hat N^i d\hat t)(d\hat x^j + \hat N^j d\hat t) ~.
\eea
We would like to write the perturbation theory in terms of the uniform inflaton gauge ($\zeta$-gauge)
\bea
\hat h_{ij}=a^2(\hat t) e^{2\zeta(\hat t,\hat \bx)} \delta_{ij} ~,~
\hat \phi(\hat t,\hat \bx) =\phi_0(\hat t) ~,~
\hat \sigma=\hat \sigma(\hat t,\hat \bx) ~,
\label{zeta_gauge}
\eea
because in this gauge the physical variables are most transparent. For single field inflation, $\zeta$ is the scalar perturbation that conserves at the superhorizon scales and directly corresponds to observables in the Cosmic Microwave background and Large Scale Structures. In our case the massive field $\hat\sigma$ is a spectator and does not play any role in reheating. Therefore the physical meaning of $\zeta$ remains the same after reheating, although as we will see it is no longer conserved outside the horizon during inflation. The $\hat\sigma$ in this gauge hence becomes the physical massive field that we will integrate over in the loops.

The simpler way to see the relevant terms in the cubic and quartic perturbative Lagrangians, especially the orders of magnitude of these terms, is however not to directly work in this gauge. In the $\zeta$-gauge, at the quartic order there are a large number of terms whose mutual cancelations are obscured by many integrations by part. Instead, it is easier to first work in the spatially flat gauge ($\delta\phi$-gauge),
\bea
h_{ij}=a^2(t) \delta_{ij} ~,~ \phi=\phi_0(t) + \delta\phi(t,\bx) ~,~
\sigma=\sigma(t,\bx) ~,
\label{phi_gauge}
\eea
and then transform it to the $\zeta$-gauge through the coordinate transformation,
\bea
\hat t = t + \delta t(t,x^i) ~, ~~~ \hat x^i = x^i + \delta x^i(t,x^i) ~.
\eea
See \cite{ChenWang} for details.
We have distinguished the two different gauges by adding hats to the variables and coordinates in the $\zeta$-gauge.
The transformation between the fields in the two gauges are given by \cite{Footnote1}
\begin{align}
\delta\phi(t,x^i) &= \dot\phi_0(t) (\delta t_{(1)} + \delta t_{(2)} + \delta t_{(3)})
\nonumber \\
&+ \half \ddot \phi_0 ( \delta t_{(1)}^2 + 2 \delta t_{(1)} \delta t_{(2)} )
+ \frac{1}{6} \dddot \phi_0 \delta t_{(1)}^3 ~,
\label{Eq:gauge_transf_deltaphi}
\\
\sigma(t,x^i) =& \hat \sigma(t,x^i) + \dot {\hat\sigma} ( \delta t_{(1)} + \delta t_{(2)} ) + \partial_i \hat\sigma \delta x^i_{(2)} + \half \ddot{\hat \sigma} \delta t_{(1)}^2 ~,
\label{Eq:gauge_transf_sigma}
\end{align}
where the subscripts in $\delta t$ and $\delta x^i$ label the perturbative orders, and we have listed terms up to the third order. For the purpose of this paper, the most important term is
\begin{align}
\delta t_{(1)} = -\zeta / H ~.
\end{align}
The orders of magnitude of the other terms are
\begin{align}
\delta t_{(2)} = \CO(\zeta^2) , ~
\delta t_{(3)} = \CO(\zeta^3, \hat\sigma^2 \zeta) ,~
\delta x_{(2)} = \CO(\zeta^2) .
\end{align}

In the slow-roll expansion, the leading order Lagrangian in the $\delta\phi$-gauge is particularly simple,
\begin{align}
  \CL_2 = \frac{a^3}{2} \delta\dot\phi^2 - \frac{a}{2} (\partial_i\delta\phi)^2
  + \frac{a^3}{2} \dot\sigma^2 - \frac{a}{2} (\partial_i\sigma)^2
  - \frac{a^3}{2} M^2 \sigma^2 ~.
  \label{Eq:CL2_delta-phi-gauge}
\end{align}
The cubic and quartic Lagrangians are of order $\CO(\sqrt{\epsilon} \delta\phi \sigma^2)$ and $\CO(\delta\phi^2 \sigma^2)$, respectively. So they are $\epsilon$-suppressed in terms of the $\zeta$-gauge.

We now apply the gauge transformation. Higher order terms generated by (\ref{Eq:gauge_transf_deltaphi}) are suppressed by $\epsilon$. So the leading terms in $\CL_3$ and $\CL_4$ in the $\zeta$-gauge are entirely generated by (\ref{Eq:gauge_transf_sigma}). Among all these terms, terms of order $\CO(\zeta\hat\sigma^2)$ generated from $\CL_2$ vanish except for temporal total derivative terms, because they are proportional to the equation of motion after integration by part.
Same for terms of order $\CO(\zeta^2\hat\sigma^2)$ generated by the third order terms in (\ref{Eq:gauge_transf_sigma}).
But the quadratic order terms in (\ref{Eq:gauge_transf_sigma}), acting on all $\sigma$'s in (\ref{Eq:CL2_delta-phi-gauge}), can generate terms of order $\CO(\zeta^2\hat\sigma^2)$.
To summarize, up to temporal total derivative terms \cite{ChenWang}, the leading terms in the $\zeta$-gauge Lagrangian are
\begin{align}
  \CL_2 &= \epsilon a^3 \dot\zeta^2 - \epsilon a (\partial_i \zeta)^2
  + \frac{a^3}{2} \dot\sigma^2 - \frac{a}{2}(\partial_i\sigma)^2 - \frac{a^3}{2} M^2 \sigma^2 ~,
  \\
  \CL_3 &= \CO(\epsilon \zeta \sigma^2) ~,   \label{Eq:CL3_zeta-gauge}
  \\
  \CL_4 &= \frac{a^3}{2H^2} \left[
  \left( \partial_t(\dot\sigma\zeta) \right)^2
  - \frac{1}{a^2} \left( \partial_i(\dot\sigma\zeta) \right)^2
  - M^2 (\dot\sigma \zeta)^2
  \label{Eq:CL4_zeta-gauge}
  \right] ~.
\end{align}
Here and below we drop all the hats for simplicity.
Each derivative on $\sigma$ produces a factor of momentum cutoff $\Lambda$ in loops. So $\CL_4$ is apparently of order $\Lambda^4$. However, using spatial integrations by part and the temporal total derivative terms, we can reduce it to order $\Lambda^2$. This avoids inconvenient cancelations between large numbers in the calculation.
Finally we perform a Legendre transform and obtain the following kinematic and leading order interaction Hamiltonian,
\begin{align}
  \CH_0 &= \epsilon a^3 \dot\zeta^2 + \epsilon a (\partial \zeta)^2
  + \frac{a^3}{2} \dot\sigma^2 + \frac{a}{2} (\partial_i \sigma)^2 + \frac{a^3}{2} M^2 \sigma^2 ~,
  \label{Eq:CH0}
  \\
  \CH_I &= - \frac{3a^3}{H} \zeta\dot\zeta \dot\sigma^2
  + \frac{a}{4} \zeta^2 (\partial_i\sigma)^2
  + \frac{9a^3}{4} M^2 \zeta^2 \sigma^2
  + \cdots ~.
  \label{Eq:CH4}
\end{align}
We have only listed three terms in $\CH_I$, which will be used in this paper as the representative examples.
A full list of terms is presented in \cite{ChenWang}.
The dots in (\ref{Eq:CH4}) include many other similar quartic order terms from (\ref{Eq:CL4_zeta-gauge}) and those generated by the Legendre transform due to $\CL_3$. The dots also include the leading cubic terms related to the temporal total derivative terms in $\CL_3$, and the precise treatment of such terms can be found in \cite{ChenWang}.

We now use the three representative terms in (\ref{Eq:CH4}) to compute the contribution of the massive field to the power spectrum.

The quantization of the fields and the mode functions followed from the kinematic Hamiltonian (\ref{Eq:CH0}) are the same as in Ref.~\cite{Chen:2012ge}. In terms of the Fourier components, we quantize
\begin{align}
  \zeta_\bk = u_\bk a_\bk + u_{-\bk}^* a_{-\bk}^\dagger ~, \quad
  \sigma_\bk = v_\bk b_\bk + v_{-\bk}^* b_{-\bk}^\dagger ~,
\end{align}
where $a_\bk$ and $b_\bk$ are independent of each other and each satisfy the usual commutation relations. The mode functions satisfy the equations of motion
\begin{align}
u_\bk''-\frac{2}{\tau} u_\bk' + k^2 u_\bk = 0 ~,
\\
v_\bk''-\frac{2}{\tau} v_\bk' + k^2 v_\bk + \frac{M^2}{H^2\tau^2} v_\bk = 0 ~,
\end{align}
where $\tau$ is the conformal time, $dt \equiv a d\tau$, and the primes denote derivatives with respect to $\tau$. The solutions are
\begin{align}
u_\bk =& -\frac{H}{\sqrt{4\epsilon k^3}} (1+i k \tau)~ e^{-ik\tau} ~,
\label{mode_function_u}
\\
v_\bk =& -i e^{i\pi/2} e^{-\pi\mu/2} \frac{\sqrt{\pi}}{2}
H (-\tau)^{3/2} H^{(1)}_{i\mu}(-k\tau) ~,
\label{mode_function_v}
\end{align}
where $\mu = \sqrt{M^2/H^2 - 9/4}$.
Both of the mode functions approach the Bunch-Davies vacuum in the short wavelength limit. The power spectrum $P_\zeta$ is defined as
\bea
\langle \zeta_{\mathbf{k}} \zeta_{\mathbf{k}'} \rangle
= \frac{P_\zeta}{2k^3} (2\pi)^5 \delta^3(\bk+\bk') ~.
\eea
At the tree level,
\bea
P_\zeta = \frac{H^2}{8\pi^2 \mpl^2 \epsilon} ~.
\label{Eq:Pzeta}
\eea
At the CMB scales $k_0=0.002{\rm Mpc}^{-1}$, $P_\zeta \approx 6.7\times 10^{-9}$.

Using the in-in formalism, the diagram (b) in Fig.~\ref{Fig:Loop_Diagrams} gives the following correction to the tree-level two-point correlation function $\langle \zeta^2 \rangle$,
\begin{align}
  &\Delta \langle \zeta^2 \rangle
  \supset 2i \left( u_{k}^*(\tau_{\rm end}) \right)^2
  (2\pi)^3 \delta^3(\bk+\bk')
  \nonumber \\
  &\times
  \int_{-\infty}^{\tau_{\rm end}} d\tau
  \left[
  -\frac{3a}{H} u_{k} u'_{k} \int \frac{d^3\bp}{(2\pi)^3} |v'_p|^2
  \right.
  \nonumber \\
  &\left. ~~
  +\frac{a^2}{4} u_{k}^2 \int \frac{d^3\bp}{(2\pi)^3} p^2 |v_p|^2
  +\frac{9M^2 a^4}{4} u_{k}^2 \int \frac{d^3\bp}{(2\pi)^3} |v_p|^2
  \right]
  \nonumber \\
  &+ {\rm c.c.} ~.
  \label{Eq:DeltaP/P}
\end{align}

The integration over the loop momentum $\bp$ is power-law divergent at UV ($p\to \infty$) and needs to be renormalized. As in particle physics, we impose a UV cutoff scale in terms of the physical momentum $\Lambda$, which in terms of the comoving momentum is $p_\Lambda = - \Lambda/(H\tau)$. This is the energy scale below which our effective field theory description is valid for {\em both} the inflaton {\em and} the massive field. The detailed value of $\Lambda$ depends on how the field theory is UV completed and will be model-dependent, and we use $\Lambda$ to parameterize our ignorance.

The integration over the conformal time $\tau$ is logarithmically divergent at IR ($\tau_{\rm end} \to 0$). This divergence has the physical interpretation that the scalar perturbation is no longer constant after the Hubble horizon exit. It receives constant contributions from the massive field loop per Hubble time. There exists a natural IR cutoff to this divergence -- the end of inflation. The result therefore is proportional to $\ln(-2k\tau_{\rm end})$, which is the number of efolds $N_e$ till the end of inflation after the mode $2k$ crosses the horizon.

\begin{figure}
\begin{center}
\includegraphics[width=8.5cm]{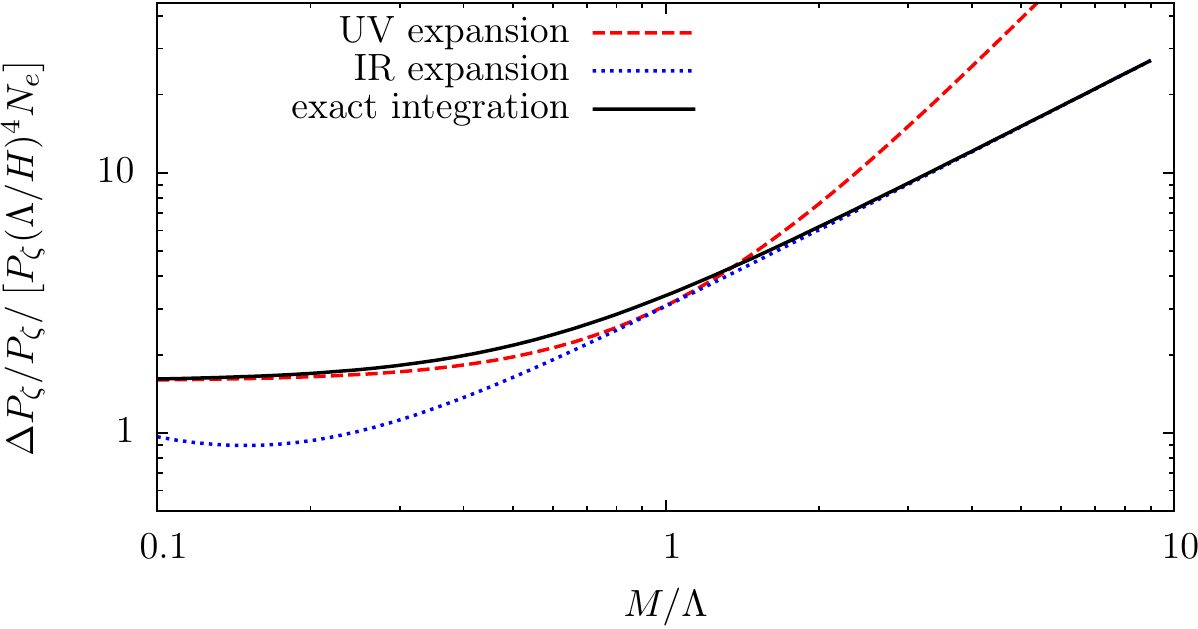}
\vspace{-0.6cm}
\end{center}
\caption{Correction (\ref{Eq:DeltaP/P}) to the curvature perturbation power spectrum as a function of isocurvaton mass, with $\Lambda=10^3 H$. }
\label{Fig:plot-correction}
\end{figure}

The integration (\ref{Eq:DeltaP/P}) can be done exactly and the results are given by special functions \cite{ChenWang}. The most important features and physical meaning of the results, however, are best illustrated by looking at some simplification limits. In the large momentum cutoff limit, $\Lambda/H \gg (M/H)^2$, we can approximate the Hankel function in $v_p$ by its UV expansion,
\begin{align}
  H^{(1)}_{i\mu}(-p\tau) \to e^{-i\pi/4} e^{\pi\mu/2} \sqrt{-\frac{2}{\pi p\tau}} e^{-ip\tau} ~.
\end{align}
From (\ref{Eq:DeltaP/P}) we get
\begin{align}
  \frac{\Delta P_\zeta}{P_\zeta} \supset
  \frac{\Lambda^4}{8\pi^2 H^2 \mpl^2 \epsilon}
  \left( \frac{19}{12} + \frac{3M^2}{2\Lambda^2} \right)
  \ln (-2k\tau_{\rm end}) ~.
  \label{Eq:UVapprox}
\end{align}
It is also interesting to look at the limit $\Lambda/H \ll \sqrt{M/H}$ in which we can approximate the Hankel function by its IR expansion,
\begin{align}
  H^{(1)}_{i\mu}(-p\tau) \to e^{i\delta} e^{\pi\mu/2}
  \sqrt{\frac{2}{\pi\mu}} (-p\tau)^{i\mu} ~,
\end{align}
where $\delta$ is an unimportant phase. In this limit,
\begin{align}
  \frac{\Delta P_\zeta}{P_\zeta} \supset
  \frac{\Lambda^4}{8\pi^2 H^2 \mpl^2 \epsilon}
  \left( \frac{3M}{\Lambda} + \frac{\Lambda}{15M} \right)
  \ln (-2k\tau_{\rm end}) ~.
  \label{Eq:IRapprox}
\end{align}
For $\Lambda\sim M$, the actual behavior is similar in order of magnitude to both expressions (\ref{Eq:UVapprox}) and (\ref{Eq:IRapprox}). In Fig.~\ref{Fig:plot-correction} we compare the exact integration with these two approximations. The other terms not listed in (\ref{Eq:CH4}) give similar behaviors with different numerical factors \cite{ChenWang}.

These results have several remarkable properties.

Firstly, comparing to the direct coupling case \cite{Kaloper:2002uj,Chen:2012ge} in which the corrections are inversely proportional to power of $M$, here $\Delta P_\zeta/P_\zeta$ is either approximately independent of $M$ or increases as $M$ increases. This is because, in the interaction Hamiltonian, the gravitational coupling is proportional to the mass-squared or momentum-squared of the massive field. At the same time, the oscillatory factor of the massive field cancel between the particle and anti-particle in the loop. Such a behavior is in sharp contradiction to the usual intuition that the massive field and high energy physics are decoupled from the low energy effective theory.

Secondly, again different from the direct coupling case, here there is no adjustable parameters one can use to turn off the coupling. Despite the Planck-mass suppression, the correction to the power spectrum is actually quite significant even if we consider single massive field. To see this, we can write (\ref{Eq:UVapprox}) as
\bea
\frac{\Delta P_\zeta}{P_\zeta} \sim P_\zeta \frac{\Lambda^4}{H^4} N_e ~,
\eea
where we have used (\ref{Eq:Pzeta}) and $\ln(-2k\tau_{\rm end}) \approx N_e\sim 60$.
For as low as $\Lambda \gtrsim M \gtrsim \CO(10^2)H$, the loop contribution from the massive field already starts to exceed the tree-level result, and the usual perturbation theory breaks down. An even stronger constraint could be obtained by examining the contribution of the logarithmic term to the tilt of the power spectrum.

Lastly, and more importantly, the results apply to all scalar fields existed during inflation \cite{Footnote2}. Among them the most interesting ones for our purpose are the massive fields with $M>H$. The existence of such fields are guaranteed in any realistic models. Furthermore, they often show up as towers of states in string theory models, with the number of states growing with the masses.

To summarize, the inflationary density perturbations turn out to be anomalously sensitive to the high energy physics, through the gravitational coupling between the inflaton and all the other states.

To further illustrate these unusual features, we apply the results to an example of extra-dimensional models. We consider $D$-dimensional spacetime compactified on a $(D-4)$-dimensional compact space, such as a $(D-4)$-torus. The sizes of the compact dimensions are of order $L$, and the fundamental Planck mass is $M_D$. The 4D Planck mass is given by
\begin{align}
\mpl^2 \sim M_D^{D-2} L^{D-4} ~.
\end{align}
The extra dimensions give rise to towers of massive states such as the scalar Kaluza-Klein states. The mass is given by the KK number $n_i$,
\bea
M_{\rm KK}^2 \sim \frac{1}{L^2} \sum_{i=1}^{D-4} n_i^2 ~.
\eea
The KK states can be described by the 4D effective field theory until the mass of the highest KK states becomes of order $M_D$, namely the KK numbers reach $\sqrt{n_i^2} = n_{\rm max} \sim M_D L$.
Any inflationary models in this setup introduce the gravitational coupling between these KK states and the inflaton. The momentum cutoff in the loop is $\Lambda \sim M_D$.
Because the density of the KK states grows with the mass, the dominant contributions come from the highest states with mass of order the cutoff scale. The total number of such states is
\bea
N_{\rm tot} \sim n_{\rm max}^{D-4}
\sim ( M_D L )^{D-4} \sim \frac{\mpl^2}{M_D^2} ~.
\eea
The net contribution of all these states is
\begin{align}
  \frac{\Delta P_\zeta}{P_\zeta} \sim
  \frac{\Lambda^4 N_e}{\epsilon H^2 \mpl^2} \times \frac{\mpl^2}{M_D^2}
  \sim \frac{M_D^2 N_e}{\epsilon H^2} ~.
\end{align}
Because $M_D >H$, it is clear that this quantity is always much larger than one. Therefore the conventional tree-level result is by no means the obvious leading order result, and it is essential to reconsider the calculation by taking into account more details of the high energy physics.

There are many important issues remaining to be investigated \cite{ChenWangWIP}. For example, it is important to understand how the density perturbation computation should work in cases where the first order gravity-mediated loop contribution becomes more important than the tree level result. One possibility is that one may be able to resum all the loop diagrams. This will completely change the prediction in the non-perturbation regime. Another possibility is to introduce supersymmetry so that the loop contribution from fermionic massive states exactly cancel the bosonic ones above the supersymmetry breaking scale, similar to what may happen to the Higgs mass hierarchy, or the cosmological constant problem. It would be interesting if supersymmetry were essential for making sense of such basic computations of density perturbations. It is important to extend the scalar in the loop to other type of fields. It is also important to consider how the calculations of the non-Gaussianities and tensor power spectrum, and how other inflation models besides the slow-roll models, are affected by the effects we considered here.

XC is supported by the Stephen Hawking Advanced Fellowship. YW is supported by a fellowship from McGill University.


\begin{thebibliography}{99}
\bibitem{Footnote1}
The tensor mode in the $\delta\phi$-gauge is ignored, since it transforms to a term that vanishes for our diagram.
\bibitem{Footnote2}
  The formalism and results we presented here can also incorporate the cases with arbitrary masses, including the massless spectator case and inflaton self-coupling \cite{ChenWang}.
\end{thebibliography}
\end{document}